\documentstyle[preprint,aps,prl]{revtex}
\begin{document}

\draft
\title{ Ab-initio theory of NMR chemical shifts in
solids and liquids}
\author{Francesco Mauri, Bernd G. Pfrommer, and Steven G. Louie}
\address{
Department of Physics, University of California at Berkeley, Berkeley, CA 94720, USA, 
and Materials Science Division, Lawrence Berkeley National Laboratory, Berkeley, CA 94720, USA.
}
\date{\today} 
\maketitle 

\begin{abstract} 
We present a theory for the {\it ab-initio} computation 
of NMR chemical shifts ($\sigma$) 
in condensed matter systems, using periodic boundary conditions.
Our approach can be applied to periodic systems such as crystals, surfaces,
or polymers and, with a super-cell technique, to non-periodic systems such as 
amorphous materials, liquids, or solids with defects.
We have computed the hydrogen $\sigma$ for a set of free molecules,
for an ionic crystal, LiH,
and for a H-bonded crystal, HF, using density
functional theory in the local density approximation. 
The results are in excellent agreement with experimental data.
\pacs{ 76.60.Cq, 71.15.Mb}
\end{abstract}

\narrowtext
Nuclear magnetic resonance (NMR) is one of the most widely used experimental
techniques in structural chemistry.
In particular, the chemical shift ($\sigma$) spectra are
a fingerprint of the molecular geometry and the chemical 
structure of the material under study.
Although the interpretation of these spectra generally relies on empirical 
rules, {\it ab-initio} calculations of $\sigma$ for molecules 
have led in many cases to  
an unambiguous determination of the microscopic 
structure\cite{strut_det_with_abinitio}.
So far,{\it  ab-initio} calculations of $\sigma$ 
have been restricted to
finite systems such as isolated molecules or 
clusters\cite{ab_initio_hf,ab_initio_dft}. 
This is a serious limitation, because  most of the NMR experiments are 
performed on liquid samples. 
Moreover, 
it is nowadays possible to measure $\sigma$ also in solids
with the resolution required for structural 
determinations\cite{enciclopedia_nmr,solid_nmr}.
E.g., $\sigma$-spectra have been used for the characterization of amorphous 
carbon\cite{nmr_amo}.  
In this letter, we present a formalism to compute, from first principles, 
$\sigma$ in extended systems using periodic boundary conditions.
Our approach can be applied to periodic systems like crystals, surfaces, 
or polymers and, 
using a super-cell technique, to non-periodic systems such as amorphous materials, 
liquids, or solids with defects. 
In the case of the amorphous solid or liquid, 
the atomic coordinates may be generated by 
{\it ab-initio} molecular dynamic 
simulations\cite{CP_water,CP_amorphousC}.

The chemical shift measures the local magnetic field in a sample induced by 
a uniform applied magnetic field.
The computation of $\sigma$ in an extended system is not straightforward,
since the expectation values of the individual terms of the 
perturbative Hamiltonian 
for extended eigenstates are not well-defined quantities\cite{zuc}.
To overcome this problem, we follow Ref.~\cite{xim}, 
in which a theory for the computation of the macroscopic 
magnetic susceptibility is presented. 
In particular: i) we obtain the magnetic 
response to a uniform field as the long-wave
limit of a periodic field; and ii) we use a generalized f-sum rule to
remove the numerical instability which occurs in this limit. 
We apply the resulting equations to real systems, describing the electronic
structure within density functional theory (DFT) in the local density 
approximation (LDA).
We compute the hydrogen $\sigma$ for a set of small molecules, for
an ionic crystal, and for a H-bonded crystal.
Our results are in excellent agreement with 
experimental data.

A uniform, external magnetic field ${\bf B}_{ext}$, applied to a 
sample induces an electronic current density ${\bf J}_{in}({\bf r})$.
This current produces an induced magnetic field ${\bf B}_{in}({\bf r})$.
If ${\bf B}_{ext}$ is small enough, a condition realized in NMR experiments, 
then:
\begin{equation}
{\bf B}_{in}({\bf r})=-{\tensor\sigma}({\bf r}){\bf B}_{ext}.
\end{equation}
Here ${\tensor \sigma}({\bf r})$ is the chemical shift tensor. 
With NMR spectroscopy, it is
possible to measure the symmetric part of 
${\tensor \sigma}({\bf r})$, or more often its trace, 
$\sigma({\bf r})=(1/3){\rm Tr} [{\tensor \sigma}({\bf r})]$, 
at the position of the non-zero spin nuclei.

In the bulk of a periodic system, 
${\tensor \sigma}({\bf r})$ is also periodic. We may write: 
\begin{equation}
{\tensor\sigma}({\bf r})= \sum_{\bf G} \tensor{\tilde{\sigma}}({\bf G}) 
e^{i{\bf G}\cdot{\bf r}},
\end{equation}
where ${\bf G}$ are the reciprocal lattice vectors.
For ${\bf G}\ne {\bf 0}$, $\tensor{\tilde{\sigma}}({\bf G})$ 
is a bulk property:
\begin{equation}
\tensor{\tilde{\sigma}}({\bf G})= -4 \pi {\tensor \chi}({\bf G},{\bf 0}),
\end{equation}
where ${\tensor \chi}({\bf G},{\bf G}')$ is the 
magnetic susceptibility matrix. However, for ${\bf G}={\bf 0}$,
$\tensor{\tilde{\sigma}}({\bf 0})$ is {\it not} a bulk property. Its value 
depends on the shape of the sample, and is determined by macroscopic
magnetostatic.
In our calculations we assume a spherical sample, for which:
\begin{equation}
\tensor{\tilde{\sigma}}({\bf 0})=-\frac{8\pi}{3}  {\tensor \chi}({\bf 0},{\bf 0}),
\end{equation}
where ${\tensor \chi}({\bf 0},{\bf 0})$ is the macroscopic 
susceptibility\cite{footnote_shape}.
Thus the calculation of $\sigma$ in a periodic system requires the knowledge 
of ${\tensor \chi}({\bf G},{\bf 0})$. 
We compute the macroscopic susceptibility 
${\tensor \chi}({\bf 0},{\bf 0})$ following Ref.~\cite{xim}.
The other elements of ${\tensor \chi}({\bf G},{\bf 0})$ are computed as
described below.

The susceptibility matrix is defined as the second derivative
of the total energy with respect to the external magnetic field. In particular,
\begin{equation}
\hat {\bf b}_{-{\bf G}}\cdot {\tensor \chi}({\bf G},{\bf 0})\hat {\bf b}_{\bf 0}=
-\left. \frac{\partial^2 E[{\bf B}] }
{\partial B_{\bf 0}\partial  B_{-{\bf G}}}\right|_{\bf B=0}=-E''_{{\bf 0},{\bf -G}}
,
\end{equation} 
where $E[{\bf B}]$ is the total energy of the system per unit volume
in the external magnetic field ${\bf B}({\bf r})$,
${\bf B}({\bf r})=[ B_{\bf 0}\hat {\bf b}_{\bf 0}+
B_{-{\bf G}} \hat {\bf b}_{-{\bf G}}\exp({-i{\bf G}\cdot{\bf r}})]$,
and $\hat {\bf b}_{-\bf G}$ and $\hat {\bf b}_{\bf 0}$ are vectors of unit 
length.
Thus 
${\tensor \chi}({\bf G},{\bf 0})$ can be evaluated using perturbation theory.
However the expectation values of 
the perturbative Hamiltonians for a uniform field
between extended eigenstates are ill defined.
To avoid this problem, we modulate the external periodic field with a finite 
wave-vector $\bf q$,  
${\bf B}({\bf r})=[ B_{\bf q}\hat {\bf b}_{\bf q}
\exp({i{\bf q}\cdot{\bf r}})+
 B_{-{\bf G-q}} 
\hat {\bf b}_{-{\bf G-q}}
\exp({-i({\bf G+q})\cdot{\bf r}})]$\cite{footnote_zero_div},
and we recover the 
results for the uniform field by considering the limit 
\cite{xim}:
\begin{equation}
\hat {\bf b}_{-\bf G}\cdot {\tensor \chi}({\bf G},{\bf 0})\hat {\bf b}_{\bf 0}= 
-\lim_
{ q \rightarrow  0} E''_{{\bf q},{\bf -G-q}}. 
\end{equation}

We now consider a spin compensated 
system described by a single particle Hamiltonian.
The derivatives of the Hamiltonian required to compute 
$ E''_{{\bf q},{\bf -G-q}}$ are:
\begin{eqnarray}
& &H'_{\bf q}= \frac{-i}{2}
\left(e^{i{\bf q}\cdot{\bf r}}{\bf a}_{\bf q}\cdot {\bf \nabla}
+ {\bf a}_{\bf q}\cdot {\bf \nabla} e^{i{\bf q}\cdot{\bf r}}\right),
\nonumber \\
& &H'_{\bf -G-q}= \frac{-i}{2}
\left(e^{-i({\bf G+q})\cdot{\bf r}}{\bf a}_{\bf -G-q} \cdot  {\bf \nabla}
\right.\nonumber \\
& & \phantom{H'_{\bf -G-q}= \frac{-i}{2}(} + \left. {\bf a}_{\bf -G-q} 
\cdot  {\bf \nabla}e^{-i({\bf G+q})\cdot{\bf r}}\right),
\nonumber \\
& & H''_{{\bf q},{\bf -G-q}}= {\bf a}_{\bf q}\cdot 
 {\bf a}_{\bf -G-q} e^{-i{\bf G}\cdot{\bf r}},\label{def_hd}
\end{eqnarray}
where ${\bf a}_{\bf s}= (i{\bf s}\times \hat {\bf b}_{\bf s})/(c s^2)$, 
and $c$ is the speed of light.
Using perturbation theory we obtain:
\begin{eqnarray}
E''_{{\bf q},{\bf -G-q}}& &= f({\bf a}_{\bf q},
{\bf a}_{\bf -G-q},{\bf G},{\bf q})\nonumber  \\
+& & 2 {\bf a}_{\bf q}\cdot
{\bf a}_{\bf -G-q} 
\int \frac{d^3 k}{(2\pi)^3}
\sum_{i\in {\cal O}} 
\langle u_{{\bf k},i}|e^{-i{\bf G}\cdot {\bf r}} |u_{{\bf k},i}\rangle,
\label{e2}
\end{eqnarray}
where 
$|u_{{\bf k},i}\rangle$ is the periodic part of the Bloch eigenstate of the
unperturbed Hamiltonian $H_{\bf k}$ with eigenvalue $\epsilon_{{\bf k},i}$,
$\cal O$ are the sets of occupied bands, and
\begin{eqnarray}
f({\bf a}_{1},{\bf a}_{2},{\bf G},{\bf q})=\int d^3r & & 
[ h({\bf a}_{1},{\bf a}_{2},{\bf r},{\bf q})
\nonumber \\
& & +h^*({\bf a}^*_{1},{\bf a}^*_{2},{\bf r},-{\bf q})]
e^{-i{\bf G}\cdot {\bf r}}, \label{def_f}
\end{eqnarray}
where the integral is performed in the periodic cell, with
\begin{eqnarray}
h({\bf a}_{1},{\bf a}_{2},{\bf r},{\bf q})&=& 
\int \frac{d^3 k}{(2\pi)^3}\sum_{i\in {\cal O}} \nonumber\\
 & & [\langle u_{{\bf k},i}|{\bf r}\rangle\langle {\bf r}|  
{\bf  a}_2 \cdot
(-i{\bf \nabla}+{\bf k}+{\bf q})|u^{{\bf q},{\bf a}_1}
_{{\bf k},i}\rangle  \nonumber \\
 & &+ \langle u_{{\bf k},i}|{\bf  a}_2 \cdot
(-i{\bf \nabla}+{\bf k})|{\bf r}\rangle\langle {\bf r}
|u^{{\bf q},{\bf a}_1}
_{{\bf k},i}\rangle ]. \label{def_h}
\end{eqnarray}
$|u^{{\bf q},{\bf a}_1}_{{\bf k},i}\rangle$ is the first order change of
the eigenstate $|u_{{\bf k},i}\rangle$ due to a field with wavevector $\bf q$.
It can be obtained by solving the linear system:
\begin{equation}
(\epsilon_{{\bf k},i}-H_{{\bf k}+{\bf q}})
|u^{{\bf q},{\bf a}_1}_{{\bf k},i}\rangle=
Q_{{\bf k}+{\bf q}}{\bf  a}_1 \cdot
\left(-i{\bf \nabla}+{\bf k}+\frac{\bf q}{2}\right) 
|u_{{\bf k},i} \rangle,\label{linear_sys}
\end{equation}
where $Q_{\bf k+q}=({\bf 1}-\sum_{i\in {\cal O}}
|u_{{\bf k+q},i}\rangle\langle u_{{\bf k+q},i}|)$
is the projector onto the empty subspace.

The first term on the r.h.s. of Eq.~(\ref{e2}) is obtained as second order 
perturbation with the 
first order derivatives of the Hamiltonian $H'_{\bf q}$ and 
$H'_{\bf -G-q}$. The second term in the r.h.s. of Eq.~(\ref{e2})
is obtained as first order perturbation with the second 
order derivative of the Hamiltonian
$H''_{{\bf q},{\bf -G-q}}$.
Since ${\bf a}_{\bf q}$ diverges as $1/q$ for $q\rightarrow 0$, the two
terms on the r.h.s. of Eq.~(\ref{e2}) individually diverge as $1/q$.
To remove this divergence, which
would produce a numerical instability, we use the generalized f-sum rule:
\begin{equation}
2 {\bf a}_{1}\cdot
{\bf a}_{2}
\int \frac{d^3 k}{(2\pi)^3}
\sum_{i\in {\cal O}}
\langle u_{{\bf k},i}|e^{-i{\bf G}\cdot {\bf r}} |u_{{\bf k},i}\rangle=-
f({\bf a}_{1},
{\bf a}_{2},{\bf G},{\bf 0}).
\label{f-sum}
\end{equation}
Substituting the f-sum rule into Eq.~(\ref{e2}) we obtain:
$ E''_{{\bf q},{\bf -G-q}}=[f({\bf a}_{\bf q},
{\bf a}_{\bf -G-q},{\bf G},{\bf q})-f({\bf a}_{\bf q},
{\bf a}_{\bf -G-q},{\bf G},{\bf 0})]$. Then,
for ${\bf G}\ne {\bf 0}$,
\begin{eqnarray}
\hat {\bf b}_{-\bf G}\cdot {\tensor \chi}& &({\bf G},{\bf 0})\hat {\bf b}_{\bf 0}
=
-\lim_{ q \rightarrow  0} E''_{{\bf q},{\bf -G-q}}\nonumber\\
=& &\left.
-\frac{\partial}{c^2G^2\partial q}
f(
\hat {\bf q}\times\hat {\bf b}_{\bf 0},
{\bf G}\times \hat{\bf b}_{-\bf G},
{\bf G}, q\hat {\bf q})\right|_{q=0},
\label{chi1}
\end{eqnarray}
where $\hat {\bf q}$ is the unit vector in the direction of $\bf q$.
Finally, 
the derivative with respect to $q$ in Eq.~(\ref{chi1}) can also be evaluated
using the following limit:
\begin{eqnarray} 
\hat {\bf b}_{-\bf G}\cdot {\tensor \chi} & &({\bf G},{\bf 0})\hat {\bf b}_{\bf 0}
= -\lim_{ q \rightarrow  0} [ 
f(\hat {\bf q}\times\hat {\bf b}_{\bf 0},
{\bf G}\times \hat{\bf b}_{-\bf G},{\bf G}, q\hat {\bf q})\nonumber \\
& & - f(\hat {\bf q}\times\hat {\bf b}_{\bf 0},
{\bf G}\times \hat{\bf b}_{-\bf G},{\bf G}, -q\hat {\bf q})]
/(2qc^2G^2).
\label{chi2}
\end{eqnarray}
Note that, for ${\bf G} \ne {\bf 0}$, ${\tensor \chi}({\bf G},{\bf 0})$
is proportional to the first derivative of $f$ with respect to 
$q$, whereas the macroscopic susceptibility
${\tensor \chi}({\bf 0},{\bf 0})$ is proportional to the
second derivative of $f$ with respect to $q$\cite{xim}.

In practice, we evaluate numerically ${\tensor \chi}({\bf G},{\bf 0})$ using Eq.~(\ref{chi2})
with a small, but finite $q$, and
the $\bf k$-integral in Eq.~(\ref{def_h}) with a finite summation
in the irreducible wedge of the Brillouin zone.

We describe the electrons using DFT-LDA; i.e., we
neglect any explicit dependence of the exchange-correlation
functional (E$_{\rm xc}$) on the current density.
The current dependence of E$_{\rm xc}$ could be taken into account
using the approximate functional proposed in Ref.~\cite{cdft},
but in practice, this produces only negligible corrections to 
$\sigma$ in real systems\cite{ab_initio_dft}.
An {\it ad-hoc} procedure to include many-body effects beyond DFT in the
the calculation of $\sigma$ has been proposed in \cite{malkinadhoc}.
While this approach improves over DFT in small molecules, 
the corrections to DFT vanish for periodic systems,
where the eigenstates are always extended.
In general, to compute the second order variation
in the DFT total energy with respect to an external perturbation,
one should take into account
the linear variation of the Hamiltonian induced by the linear
variation of the charge $\delta \rho$.
However, if the perturbation is a magnetic field, $\delta \rho$ is zero
by time reversal symmetry. Thus Eq.s~(\ref{def_hd}-\ref{chi2}) are correct
within DFT.
  
In the present calculation, we will consider the magnetic response 
of valence electrons only.
We describe the ionic cores by norm conserving 
pseudopotentials\cite{tm2} in the Kleinman-Bylander form\cite{kb}.
This approximation does not affect $\sigma$ of the nuclei without
core electrons, such as H\cite{ab_initio_hf,ab_initio_pseudo}, but
those containing core electrons.
In the latter case $\sigma$ computed with pseudopotentials differs 
from the one computed with an all-electron scheme by three different
terms:
i) the diamagnetic core contribution, which is independent 
of the chemical environment;
ii) a contribution due 
to the transitions from valence states to core states\cite{xim};
iii) a contribution due to the difference between the all-electron valence
wavefunctions and the pseudo wavefunctions in the core region.
We found that, for first row atoms like carbon, 
the error due to the pseudopotential is minor, 
since the terms ii) and iii) are usually much smaller 
than the range of variation of $\sigma$ with the chemical environment.
In the present paper, however, we present only the results for 
$\sigma$ of H, which is not affected by the use of pseudopotentials. 
Finally, in our pseudopotential calculation, we replace 
$(-i{\bf \nabla} +{\bf k}+{\bf q}/2)$ in Eq.~(\ref{linear_sys}) with
$({\bf v}^p_{\bf k}+{\bf v}^p_{\bf k+q})/2$, where 
${\bf v}^p_{\bf k} = (d/d{\bf k})H^p_{{\bf k}}$ is the velocity operator,
and $H^p_{{\bf k}}$ is the pseudo-Hamiltonian\cite{footnote_pseudo}.

We expand the wave-functions using a plane wave (PW) basis set.
We obtain $|u^{{\bf q},{\bf a}_1}_{{\bf k},i}\rangle$
by solving Eq.~(\ref{linear_sys}) with a conjugate gradient 
minimization\cite{num_rec}; i.e., we never use the unoccupied eigenstates
of $H^p_{\bf k}$.
We obtain $f({\bf a}_{1},{\bf a}_{2},{\bf G},{\bf q})$  
from $|u^{{\bf q},{\bf a}_1}_{{\bf k},i}\rangle$ using fast Fourier transforms.
Thus, the computational effort to evaluate
$\sigma$ for all nuclei in the sample is comparable to that required
for a computation of the total energy\cite{scaling}.
We compute $\sigma$ in systems containing atoms of H, C, Li, and F.
For H we use a local pseudopotential to speed up the convergence with respect
to the PW basis\cite{footnote_h}.
For C, Li and F atoms we use a pseudopotential with a non-local s-projector, 
to describe the interaction
of valence states with the 1s core states.
We use a PW energy cut-off of 100 Ry 
for the F atom, and of 70 Ry for the other atoms.
Molecular and crystalline structures are taken from experiments
\cite{geometries}. 
The results for the free molecules are obtained using super-cells,
such that the error on $\sigma$ due to interaction between
a molecule and its periodic replica is smaller than 0.1 parts 
per million (ppm).

The results for a set of molecules are reported in 
Table~\ref{moleculetable}.
All the computed values are in excellent agreement with experimental data.
The discrepancies are larger for C$_2$H$_4$ and C$_2$H$_2$, in which a double
and a triple C-C bond exist, respectively. This could indicate a relatively 
larger inaccuracy of LDA in describing these types of bonds. 
Similar agreement with experiments for the hydrogen $\sigma$ in molecules 
is found in previous calculations\cite{ab_initio_hf,ab_initio_dft}.

The results for molecular and crystalline LiH are reported in
Table~\ref{lihtable}.
LiH is an ionic system, which
crystallizes in the rock-salt structure.
The Li-H equilibrium distance in the crystal
is significantly larger than that of the molecule.
Our results show that the difference between $\sigma$ in the molecule 
and in the solid is very small.
To verify if $\sigma$ is sensitive to the geometry,
we consider a crystal at a pressure of 65~GPa, in which
the Li-H distance is equal to that  of the molecule.
In this case we find a much larger value of $\sigma$.
Thus, $\sigma$ of the crystal and of the free molecule are only similar 
at the equilibrium geometries.

The results for molecular and crystalline HF are reported in
Table~\ref{hftable}.
The intra-molecular bond of HF is covalent.
In the liquid and solid phases, the HF molecules are
bonded together in zigzag chains via H-bonds.
HF has the strongest H-bonding found in nature.
This strong inter-molecular bond is reflected in the large 
variation of $\sigma$ observed in the transition from the gas phase to the
liquid phase, and in the large temperature dependence
of $\sigma$ in the liquid phase\cite{sigmaexp2}.
The HF crystal is constituted of a 2D stacking of parallel 1D zigzag chains.
We compute $\sigma$ for the crystal at the experimental geometries
measured for a DF crystal at two temperatures, 4.2K and 
85K\cite{hfstructure}.
The computed $\sigma$ in the free HF molecule is in outstanding agreement
with experimental data.
In the crystal, the computed $\sigma$ decreases dramatically to a value close 
to the one measured in the liquid phase (we are not aware of any experimental 
measurement of $\sigma$ for solid HF).
Also the theoretical variation of $\sigma$ with temperature, 
$\delta\sigma/\delta T=0.013\,{\rm ppm/K}$, is
very close to the one measured in the liquid, which is
$d\sigma/dT=0.0135\,{\rm ppm/K}$\cite{sigmaexp2}.
Thus, with respect to $\sigma$, the behavior of the solid 
is very similar to that of the liquid.
Interestingly, a single H-bond does not account for the large decrease
of $\sigma$ observed in the condensed phases.
Indeed the computed $\sigma$ for the central H atom in a HF-HF dimer, is
closer to the $\sigma$ of the free molecule than to that of the 
liquid\cite{ab_initio_hf}.
 
In conclusion, we have presented an {\it ab-initio} theory for 
the evaluation of NMR chemical shifts ($\sigma$) in extended systems.
We have computed, for the first time, $\sigma$ in real solids.
Our results show that DFT-LDA predicts hydrogen $\sigma$
which are in excellent agreement with experiment.
The evaluation of $\sigma$ 
requires the same numerical effort as
the computation of the total energy\cite{scaling}.
Thus, our approach can be applied to super-cells containing 
up to a hundred atoms, 
which are sufficient to model liquids or amorphous 
materials\cite{nmr_amo,CP_water,CP_amorphousC}.

We thank Prof. U. Haeberlen and Prof. M. Mehring for useful discussions.
This work was supported by the National Science
Foundation under Grant No. DMR-9520554, 
by the Office of Energy Research, Office of
Basic Energy Sciences, 
Materials Sciences Division of the U.S. Department of Energy
under Contract No. DE-AC03-76SF00098,
and by the Miller Institute for Basic Research in Science.
Computer time was provided by the NSF at the NCSA in Illinois and by NERSC 
at LBNL.

\begin{table}
\begin{tabular}{lcc}
&$\sigma$-theory (ppm)& $\sigma$-experiment (ppm)\\
\tableline
H$_2$        &25.9 &26.2$^a$\phantom{1}\\
CH$_4$       &30.7 &30.61$^a$\\
C$_2$H$_6$   &29.7 &29.86$^b$\\
C$_2$H$_4$   &24.5 &25.43$^b$\\
C$_2$H$_2$   &28.6 &29.26$^b$
\end{tabular}
\caption{
\label{moleculetable}
Computed and measured H chemical shifts ($\sigma$) for a set of free 
molecules.$^a$  Ref.\protect\cite{sigmaexp1},$^b$ Ref.\protect\cite{sigmaexp2}.}
\end{table}

\begin{table}
\begin{tabular}{llc}
            &           & $\sigma$-theory (ppm)\\
\tableline
LiH molecule& $d_0=1.595$&  26.6\\
LiH crystal & $a=a_0=4.08$ &  26.3\\
LiH crystal & $a=2d_0=3.19$ &  31.2\\
\end{tabular}
\caption{
\label{lihtable}
Computed H chemical shifts ($\sigma$) in LiH for a free molecule and a 
spherical crystalline sample (in the rock-salt structure).
$d_0$ is the equilibrium bond-length of the molecule,
$a$ is the lattice constant, and $a_0$ is the experimental
equilibrium lattice constant.
For $a=2d_0$, 
the shortest Li-H distance in the crystal is equal to $d_0$.
}
\end{table}

\begin{table}
\begin{tabular}{lccc}
             &     T (K)         & $\sigma$-theory (ppm)&$\sigma$-experiment (ppm)\\
\tableline
HF molecule&     & 28.4 & 28.5$^a$ \\
HF crystal & 4.2 & 20.71 &      \\
HF crystal & 85  & 21.76 &      \\
HF liquid  & 214 &      & 21.45$^b$\\
\end{tabular}
\caption{
\label{hftable}
Computed and measured hydrogen chemical shifts $\sigma$ in HF for a free molecule, a
spherical crystalline sample, and a spherical liquid sample. T is the temperature.
$^a$  Ref.\protect\cite{sigmaexp1},$^b$ Ref.\protect\cite{sigmaexp2}.}
\end{table}
 

\begin{references}
\bibitem{scaling}
The computational time required for the evaluation of all $\sigma$ in a unit 
cell scales with the system size as  
$O(N_{\cal O}^2N_{pw})$, where $N_{\cal O}$ is the number of
occupied states and $N_{pw}$ is the number of PWs.
\bibitem{footnote_h}
We verified that the values for H $\sigma$ obtained with the H pseudopotential are rigidly
shifted by -0.3 ppm with respect to the values obtained without the H pseudopotential.
The results reported in the present paper have been corrected for this effect.
\end{references}
\end{document}